# Inherent piezomagnetic, piezoelectric, linear magnetoelectric effects and built-in fields in nanos

E.A. Eliseev,[1] A.N. Morozovska,[2,*], M.D. Glinchuk[1,†] B.Y. Zaulychny[1], V.V. Skorokhod[1], and R. Blinc[3]

[1] Institute for Problems of Materials Science, NAS of Ukraine,
Krjijanovskogo 3, 03142 Kiev, Ukraine,

[2] V. Lashkarev Institute of Semiconductor Physics, NAS of Ukraine,
prospect Nauki 41, 03028 Kiev, Ukraine

[3] Jožef Stefan Institute, P. O. Box 3000, 1001 Ljubljana, Slovenia

**Abstract**

The symmetry breaking inevitably present in the vicinity of any surface, namely an inversion center disappears in surface normal direction and only axes and planes normal to the surface conserve, gives rise to the spontaneous piezomagnetic, piezoelectric and magnetoelectric effects in nanosystems, while the effects can be absent in a bulk material. All these phenomena are inherent to nanos made of materials belonging to all ninety bulk magnetic classes. Therefore the new linear magnetoelectrics should appear among nanomaterials, nonpiezomagnetic and nonpiezoelectric in the bulk. To demonstrate this we consider the typical cases of ultra-thin films, nanowires and nanospheres.

Coupled with a surface stress for nanoparticles and a mismatch strains for thin films on substrates the surface piezomagnetic and piezoelectric effects lead to the appearance of built-in magnetic and electric fields respectively. The built-in fields play an important role in the appearance of self-magnetization and self-polarization in the nanosystems paramagnetic and paraelectric in the bulk and can lead to the appearance of some other interesting properties absent in the bulk.

We obtained analytical dependencies on sizes for the built-in fields and magnetoelectric coupling coefficients. The values of the built-in fields increase with the decrease of film thickness $h$ or nanoparticles radii $R$ as $\sim 1/h$ or $1/R^2$ respectively, while the magnetoelectric coupling is inversely proportional to the sizes in both cases. This shows the strong influence of sizes on the considered properties of nanos and so opens the ways to govern the properties by the choice of the sizes and to create new multifunctional nanomaterials.

The appearance in nanos of new terms in optic, galvanic and thermagnetic effects are forecasted.

---

[*] Corresponding author: morozo@i.com.ua
[†] Corresponding author: glin@ipms.kiev.ua

# I. Introduction

The properties of nanos absent in the bulk are the most interesting [1, 2, 3, 4, 5, 6]. Such striking phenomenon as the observation of ferromagnetism in spherical nanoparticles (size 7–30 nm) of nonmagnetic oxides such as $CeO_2$, $Al_2O_3$, ZnO, etc has been reported in the paper [1]. Extremely strong superparamagnetic behavior down to 4 K has been found in gold and palladium nanoparticles with mean diameter 2,5 nm and narrow sizes distribution with no magnetization in bulk [2]. Ferroelectric phase transition appears in thin antiferroelecric $PbZrO_3$ and $BiNbO_4$ films under their thickness decrease [3, 7]. It is interesting that iron that is a typical ferromagnetic material turns antiferromagnetic in a monolayer on W(001) [6]. The strong enhancement of spontaneous polarization and ferroelectric phase conservation up to the chemical decomposition has been observed in Rochelle salt nanorods of diameter 30 nm [4]. The appearance of ferroelectricity takes place in nanorods and thin films of the incipient ferroelectrics, which remain paraelectric up to zero K in the bulk [5, 8, 9]. Given examples illustrate how dramatically different from bulk behaviour materials can be at low dimensions.

It was shown in the papers [11-9] that the possible physical origin of polarization and magnetization enhancement could be mechanical conditions in restricted geometry of nanomaterials and in particular the built-in magnetic and electric fields originated from piezoeffects and e.g. surface tension. While the existence of piezoelectric effect in the vicinity of surface was shown earlier [10, 11], the appearance of piezomagnetic effect was the assumption of [12, 13] that up to now was not approved for nanosystems. The same reasons were shown to be the source of intriguing modifications of the structure polar, magnetic and electronic state [14, 15, 16], electret state appearance and ferroelectric thin films self-polarization that was widely used as important technological process, but stayed unexplained for many years up to 2004 [10].

The magnetoelectric (ME) effect, i.e. when the application of either a magnetic field or an electric field induces an electric polarization as well as magnetization, attracted much attention in the last years [17, 18, 19, 20]. The importance of linear ME effect for the applications in modern technology and in particular for memory devices increases essentially the interest to magnetoelectric materials. Recently the revival of the ME effect has been observed due to discovery of high (several hundred percents) ME effects both in single phase and composite materials [21, 22, 23, 24]. However up to now the efforts of scientists and engineers are directed to the search for new materials with high ME effect.

In this paper we will show that for nanomaterials both with phase transitions (ferroics) and without any phase transitions like above-mentioned oxides, which belong to all ninety magnetic classes, piezomagnetic, piezoelectric, linear ME effect and built-in magnetic and electric fields are inherent characteristics. The main advantage of nanos is the possibility to govern their properties by the choice of their geometry and sizes as illustrated by analytical expressions derived in the paper. They open the way to obtain necessary characteristics and govern the considered and other properties of nanos by the choice of their sizes and surface geometry.

**II. Piezomagnetic, piezoelectric and magnetoelectric effects**

A piezomagnetic tensor, that coupled the axial vector of magnetization **M** with the polar tensor of strain $u_{ij}$, is the third rank axial tensor $d_{ijk}^{(m)}$. In bulk materials the piezomagnetic effect could exist in 66 magnetic classes [25]. The number was obtained as a difference between the total number of magnetic classes (90) and the number (21) of magnetic classes, which possess operations of time reversal and space inversion center simultaneously [25]. Because the strain tensor is symmetric ($u_{ij} = u_{ji}$), the tensor $d_{ijk}^{(m)}$ is equal to zero for the three classes of cubic symmetry: 43m, 432 and m3m, which have to be excluded also.

It is pretty obvious that because the inversion center is absent in the vicinity of surface for nanos of arbitrary geometry, piezomagnetic effect has to exist in the aforementioned 21 magnetic classes. The calculations have shown that the piezomagnetic effect exists in abovementioned 3 cubic classes also. Therefore, contrary to the bulk, in nanos piezomagnetic effect has to exist in 90 bulk magnetic classes.

That components of any tensor is defined by its transformation law, lets consider quantitatively the form of piezomagnetic tensor presentation in nanos. To find out the nonzero components of third rank tensors we will use the system of linear equations obtained from the transformation laws for the axial ($d_{lpn}^{(m)}$) and polar ($d_{lmn}^{(e)}$) third rank tensors describing piezomagnetic (*m*) and piezoelectric (*e*) effects in bulk [25]:

$$\tilde{d}_{ijk}^{(m)} = (-1)^{tr} \det(\mathbf{A}) A_{il} A_{jp} A_{kn} d_{lpn}^{(m)}, \qquad \tilde{d}_{ijk}^{(e)} = A_{il} A_{jm} A_{kn} d_{lmn}^{(e)}. \qquad (1)$$

Hereinafter the summation is performed over the repeating indexes. **A** is the transformation matrix with components $A_{ij}$ ($i,j = 1,2,3$) and determinant $\det(\mathbf{A}) = \pm 1$; the factor *tr* denotes either the presence (*tr* = 1) or the absence (*tr* = 0) of time-reversal operation coupled to the entirely space point transformation $A_{ij}$. For the case when the matrices **A** represent all the generating elements of the material point symmetry group (considered hereinafter) the identity $\tilde{d}_{ijk}^{(m,e)} \equiv d_{ijk}^{(m,e)}$ should be valid for nonzero components of the piezotensors.

For any spatially confined system the inversion center disappears in surface normal direction and only the symmetry axes and planes normal to the surface conserves [26]. Thus the magnetic and space symmetry groups should be changed as the new transformation matrices $A_{ij}^S$ is the sub-group of initial point symmetry group $A_{ij}$, namely rotations and mirror reflections have the form

$$\mathbf{A}^S = \begin{pmatrix} a_{11} & a_{12} & 0 \\ a_{21} & a_{22} & 0 \\ 0 & 0 & 1 \end{pmatrix}$$ for the surface normal $\uparrow\uparrow x_3$, at that $\det(\mathbf{A}^S) = \det(\mathbf{A}) = \pm 1$. As a result the surface piezoeffect tensors $d_{lpn}^{(Sm)}$ and $d_{lmn}^{(Se)}$ (existing even in a cubic symmetry lattice near the surface) should obey another transformation laws than the ones existing in the bulk of material, namely $d_{ijk}^{(Sm)} \equiv (-1)^{tr} \det(\mathbf{A}^S) A_{il}^S A_{jp}^S A_{kn}^S d_{lpn}^{(Sm)}$ and $d_{ijk}^{(Se)} \equiv A_{il}^S A_{jm}^S A_{kn}^S d_{lmn}^{(Se)}$.

By the same way one can analyze the second rank ME tensor $\gamma_{ij}$. The transformation laws for the surface linear ME effect axial tensor $\gamma_{ij}^S$ are $\tilde{\gamma}_{ij}^S = (-1)^{tr} \det(\mathbf{A}^S) A_{ik}^S A_{il}^S \gamma_{kl}^S$ and $\tilde{\gamma}_{ij}^S \equiv \gamma_{ij}^S$ for nonzero components, i.e. the laws differ from the one existing in the bulk of material: $\tilde{\gamma}_{ij} = (-1)^{tr} \det(\mathbf{A}) A_{ik} A_{il} \gamma_{kl}$. In bulk of materials the ME effect was shown to exist in 58 magnetic classes [25]. The analysis, similar to the one we performed for piezomagnetic tensor, shows that ME effect exists in nanos in ninety magnetic classes. Thus the new piezomagnetics, piezoelectrics and linear magnetoelectrics should appear even among nanomaterials nonpiezomagnetic and nonpiezoelectric in the bulk, e.g. simple binary oxides like EuO and CoO.

To demonstrate the intriguing possibility, we calculate the evident form of the surface piezoelectric, piezomagnetic and ME tensors for the bulk m3m, m′3m′, m′3m and m3m′ cubic symmetry groups (symbol prime stands for the coupling with time reversal). The surface 4mm, 4m′m′, 4′m′m and 4′mm′ symmetry groups were directly obtained from the bulk m3m, m′3m′, m′3m and m3m′ symmetry groups respectively by adding the surface with normal $x_3 \uparrow\uparrow 4$ (symbol 4 stands for the forth order rotation axis). Note, that the m3m, m′3m, m3m′, m′3m′ symmetry groups correspond to the bulk symmetry of the nonpiezoelectric binary oxides MnO, FeO, CoO, NiO, MnS, EuO, PrO and the paraphrase of the BiFeO$_3$.

Results for $d_{ijk}^{(Sm)}$, $d_{ijk}^{(Se)}$ and $\gamma_{ij}^S$ are presented in the Table 1. It is seen from the Table 1 that a nonpiezoelectric and nonpiezomagnetic bulk material with m3m, m′3m′, m′3m and m3m′ symmetry becomes piezoelectric and piezomagnetic in the vicinity of surface with different $d_{ijk}^{(Sm)}$ tensors, which depend on the surface symmetry group, the influence of surface on the symmetry and properties being essential on the distances of several tens nm from the surface [26, 27, 28]. Nonpiezoelectric but piezomagnetic bulk materials with m3m′ symmetry remain piezomagnetic and become piezoelectric in the vicinity of surface, but the symmetry of $d_{ijk}^{(Sm)}$ changes in comparison with a bulk tensor $d_{ijk}^{(m)}$. It is seen from the Table 1 that the choice of surface influences the number and type of nonzero components of piezoelectric, piezomagnetic and ME tensors.

Nonmagnetoelectric bulk materials with m3m, m′3m and m3m′ symmetry become linear magnetoelectric in the vicinity of surface with different $\gamma_{ij}^S$ tensors, which depend on the surface symmetry group. Magnetoelectric bulk materials with m′3m′ symmetry remain linear magnetoelectrics in the vicinity of surface, but the symmetry of $\gamma_{ij}^S$ changes in comparison with a bulk tensor $\gamma_{ij}$ (see the last row in the Table 1).

Thus the symmetry breaking inevitably present in the vicinity of the any surface gives rise to the piezomagnetic, piezoelectric and ME effects in nanosystems, while the effects can be absent in a bulk material. Linear ME effect can exist in nanos made of materials belonging to the all ninety bulk magnetic classes. We have to underline, that this number is much larger than the bulk magnetic classes number 66, as well as the magnetoelectric classes number 58.

**Table 1.** Surface and bulk piezoelectric, piezomagnetic and ME tensors.

| Symmetry group | Piezomagnetic tensor non-trivial components | Piezoelectric tensor non-trivial components | Linear ME tensor non-trivial components |
|---|---|---|---|
| **Bulk** m3m, m′3m | Absent in the bulk $d_{ijk}^{(m)} \equiv 0$ | Absent in the bulk $d_{ijk}^{(e)} \equiv 0$ | Absent $\gamma_{ij}=0$ |
| **Bulk** m′3m′ | Absent in the bulk $d_{ijk}^{(m)} \equiv 0$ | Absent in the bulk $d_{ijk}^{(e)} \equiv 0$ | $\begin{pmatrix} \gamma_{11} & 0 & 0 \\ 0 & \gamma_{11} & 0 \\ 0 & 0 & \gamma_{11} \end{pmatrix}$ |
| **Bulk** m3m′ | $d_{123} = d_{213} = d_{312} = d_{14}^{(m)}$ $\begin{pmatrix} 0 & 0 & 0 & d_{14}^{(m)} & 0 & 0 \\ 0 & 0 & 0 & 0 & d_{14}^{(m)} & 0 \\ 0 & 0 & 0 & 0 & 0 & d_{14}^{(m)} \end{pmatrix}$ | Absent in the bulk $d_{ijk}^{(e)} \equiv 0$ | Absent $\gamma_{ij}=0$ |
| **Surface** 4mm (normal $x_3 \uparrow\uparrow 4$) | $d_{14}^{(Sm)} = d_{123} = -d_{213}$ $\begin{pmatrix} 0 & 0 & 0 & d_{14}^{(Sm)} & 0 & 0 \\ 0 & 0 & 0 & 0 & -d_{14}^{(Sm)} & 0 \\ 0 & 0 & 0 & 0 & 0 & 0 \end{pmatrix}$ | $d_{31}^{(Se)}=d_{311}=d_{322}, d_{333}=d_{33}^{(Se)}$, $d_{131} = d_{232}= d_{15}^{(Se)}$ $\begin{pmatrix} 0 & 0 & 0 & 0 & d_{15}^{(Se)} & 0 \\ 0 & 0 & 0 & d_{15}^{(Se)} & 0 & 0 \\ d_{31}^{(Se)} & d_{31}^{(Se)} & d_{33}^{(Se)} & 0 & 0 & 0 \end{pmatrix}$ | $\gamma_{12}^S = -\gamma_{21}^S$ $\begin{pmatrix} 0 & \gamma_{12}^S & 0 \\ -\gamma_{12}^S & 0 & 0 \\ 0 & 0 & 0 \end{pmatrix}$ |
| **Surface** 4m′m′ (normal $x_3 \uparrow\uparrow 4$) | $d_{311}=d_{312}=d_{31}^{(Sm)}, d_{333}=d_{33}^{(Sm)}$, $d_{131} = d_{232}= d_{15}^{(Sm)}$ $\begin{pmatrix} 0 & 0 & 0 & 0 & d_{15}^{(Sm)} & 0 \\ 0 & 0 & 0 & d_{15}^{(Sm)} & 0 & 0 \\ d_{31}^{(Sm)} & d_{31}^{(Sm)} & d_{33}^{(Sm)} & 0 & 0 & 0 \end{pmatrix}$ | The same as above $d_{31}^{(Se)}=d_{311}=d_{322}, d_{333}=d_{33}^{(Se)}$, $d_{131} = d_{232}= d_{15}^{(Se)}$ | $\gamma_{11}^S = \gamma_{22}^S$ $\begin{pmatrix} \gamma_{11}^S & 0 & 0 \\ 0 & \gamma_{11}^S & 0 \\ 0 & 0 & \gamma_{33}^S \end{pmatrix}$ |
| **Surface** 4′mm′ (normal $x_3 \uparrow\uparrow 4$, $x_{1,2} \perp$ m-planes)* | $d_{123} = d_{213} = d_{14}^{(Sm)}, d_{312} = d_{36}^{(Sm)}$ $\begin{pmatrix} 0 & 0 & 0 & d_{14}^{(Sm)} & 0 & 0 \\ 0 & 0 & 0 & 0 & d_{14}^{(Sm)} & 0 \\ 0 & 0 & 0 & 0 & 0 & d_{36}^{(Sm)} \end{pmatrix}$ | The same as above $d_{31}^{(Se)}=d_{311}=d_{322}, d_{333}=d_{33}^{(Se)}$, $d_{131} = d_{232}= d_{15}^{(Se)}$ | $\gamma_{12}^S = \gamma_{21}^S$ $\begin{pmatrix} 0 & \gamma_{12}^S & 0 \\ \gamma_{12}^S & 0 & 0 \\ 0 & 0 & 0 \end{pmatrix}$ |
| **Surface** 4′m′m (normal $x_3 \uparrow\uparrow 4$, $x_{1,2} \perp$ m′-planes)* | $d_{113} = -d_{223} = d_{15}^{(Sm)}$, $d_{322} = -d_{311} = d_{31}^{(Sm)}$ $\begin{pmatrix} 0 & 0 & 0 & 0 & d_{15}^{(Sm)} & 0 \\ 0 & 0 & 0 & -d_{15}^{(Sm)} & 0 & 0 \\ d_{31}^{(Sm)} & -d_{31}^{(Sm)} & 0 & 0 & 0 & 0 \end{pmatrix}$ | The same as above $d_{31}^{(Se)}=d_{311}=d_{322}, d_{333}=d_{33}^{(Se)}$, $d_{131} = d_{232}= d_{15}^{(Se)}$ | $\gamma_{11}^S = -\gamma_{22}^S$ $\begin{pmatrix} \gamma_{11}^S & 0 & 0 \\ 0 & -\gamma_{11}^S & 0 \\ 0 & 0 & 0 \end{pmatrix}$ |

*)Groups 4′mm′ and 4′m′m are equivalent within the rotation of the coordinate system

The proposed method of piezomagnetic, piezoelectric and ME tensors nonzero components calculations for nanos of different geometry was applied to all 90 bulk magnetic groups and obtained results will be published elsewhere.

### III. Built-in magnetic and electric fields

Below we demonstrate that the surface piezomagnetic and piezoelectric effects coupled with the surface stress for nanoparticles and a mismatch strains for thin films on substrates lead to the appearance of built-in magnetic and electric fields (the latter were calculated earlier for ferroics [10, 11]). The surface stress or strains are strongly dependent on the boundary conditions and the ambient material or substrate [29, 30].

The built-in fields play an important role in the appearance of self-magnetization and self-polarization in the nanos paramagnetic and paraelectric in the bulk [12].

For correct phenomenological description of any confined (and in particular nanosized) system the **surface energy $G_S$** should be considered, as its contribution in the free energy $G = \int_S g_S d^2 r + \int_V g_V d^3 r$ increases with the system size decrease, i.e. it increases with $S/V$ ratio increase ($S$ is the system surface, and $V$ is its volume). Following [17], the excess of the free energy density related with the electric and magnetic field components $E_i$ and $H_i$, and the stress tensor $\sigma_{ij}$ has the form:

$$g_S = -\left(d_{ijk}^{(Se)}\sigma_{jk}E_i + d_{ijk}^{(Sm)}\sigma_{jk}H_i + \gamma_{ij}^S H_i E_j\right), \tag{2a}$$

$$g_V = -\left(\begin{array}{l}\left(P_{0i} + \dfrac{\varepsilon_0}{2}\chi_{ij}^e E_j\right)E_i + \left(M_{0i} + \dfrac{\mu_0}{2}\chi_{ij}^m H_j\right)H_i \\ + d_{ijk}^{(e)}\sigma_{jk}E_i + d_{ijk}^{(m)}\sigma_{jk}H_i + \gamma_{ij}H_i E_j + \dfrac{s_{jklm}}{2}\sigma_{jk}\sigma_{lm} + ...\end{array}\right). \tag{2b}$$

$s_{ijkl}$ is the elastic compliances tensor. $P_{0i}$ and $M_{0i}$ are the spontaneous polarization and magnetization vector components, $\chi_{ij}^e$ and $\chi_{ij}^m$ are electric and magnetic susceptibilities, $\varepsilon_0$ and $\mu_0$ are universal dielectric and magnetic constants respectively.

All higher nonlinear terms and the gradient energy were omitted in Eqs.(2) for the sake of simplicity. Note, that hereinafter we will not consider metallic nanos, since electrical polarization is absent in them, but it is not excluded that for metallic materials Eqs. (2) could be valid without piezoelectric effect and polarization.

Minimization of the free energy (2) with respect to the electric, magnetic fields and stress components leads to the equations of state for the polarization $P_i = -(\partial G/\partial E_i)$, magnetization $M_i = -(\partial G/\partial H_i)$ and strain tensor components $u_{ij} = -(\partial G/\partial \sigma_{ij})$. Rigorous methods for the solution of these equations with respect to boundary conditions for the stress or strains, and the variation of the surface energy are described in details in Refs.[12, 8]. Both rigorous methods and simplified core-and-shell approach presented in the Appendix lead to the renormalized free energy density: $g_R \sim -\left(d_{ijk}^{(e)}\sigma_{jk}E_i + d_{ijk}^{(m)}\sigma_{jk}H_i + \varepsilon_0 E_i^b E_i + \mu_0 H_i^b H_i + \gamma_{ij}^R H_i E_j\right)$. It includes the magnetic ($H_i^b \cong \frac{S}{\mu_0 V} d_{ijk}^{(Sm)} \sigma_{jk}$) and electric ($E_i^b \cong \frac{S}{\varepsilon_0 V} d_{ijk}^{(Se)} \sigma_{jk}$) built-in fields and magnetoelectric energy density $g_{ME} = \gamma_{ij}^R H_i E_j$ with renormalized ME coefficient $\gamma_{ij}^R \cong \gamma_{ij} + \frac{S}{V} \gamma_{ij}^S$, where $\gamma_{ij}^S$ is listed in the Table 1.

General expressions for the built-in fields $H_i^b$ and $E_i^b$ produced by the surface 4mm, 4m′m′, 4′mm′ and 4′m′m symmetry groups (obtained from the bulk m3m, m′3m, m3m′ and m′3m′ symmetry groups respectively) and nonzero ME coupling coefficients $\gamma_{ij}^R$ size dependences are listed in the Table 2 for thin films and nanoparticles of different geometry.

It is seen from the table, that the built-in fields and linear magnetoelectric coupling $\gamma_{ij}^R$ spontaneously arises for the typical cases of ultra-thin films, nanowires, nanotubes and nanospheres. So, the large amount of new linear magnetoelectrics should appear among nanos, which are nonmagnetoelectric in the bulk.

It is necessary to underline that the values of the built-in fields and ME coupling increase with the decrease of film thickness $h$ or nanoparticles radii $R$. Obtained analytical dependencies for thin films or nanoparticles respectively have shown that $H_i^b$ and $E_i^b$ ~$1/h$ or $1/R^2$, while the ME coupling $\gamma_{ij}^R$ is inversely proportional to the sizes in both cases. This shows the strong influence of sizes on the nanos properties and so opens the ways to govern the considered properties by the choice of the sizes. In particular the linear ME coupling should dramatically changes the phase diagrams of ferroic nanosystems with various geometry.

Inevitable symmetry breaking in the vicinity of surface (and thus in thin films and small enough nanoparticles) could lead to new terms in optical properties of nanos including linear electrooptical and magnetooptical effects and in nonlinear susceptibilities. In particular, keeping in mind that electrooptical effect and nonlinear susceptibilities are defined by the third rank tensor, the results presented in the first and second columns of the Table 1 can be applied for these effects also. New terms in galvanic and thermomagnetic effects namely new nonzero Hall, Rigi, Leduke, Nernst, Ettingaus and magnetoresistance coefficients for nanos could appear and can be calculated by the same way we proposed here. The detailed calculations of these effects are in progress now.

**Table 2.** Surface built-in fields and ME coupling coefficients induced by the surface piezoelectric and piezomagnetic effects (see Appendix for details).

| Nano-system | Built-in magnetic field normal component H | Built-in electric field normal component E | Linear ME coupling* $\gamma_{ij}^R = \gamma_{ij} + \dfrac{\gamma_{ij}^S}{\eta}$, $\eta$ is the characteristic size |
|---|---|---|---|
| Thin layer of thickness $h$, surface normal $\uparrow\uparrow x_3$ $\sigma_{ij}$ are the surface stress tensor | $H_1^b = \dfrac{d_{113}^{(Sm)}\sigma_{13} + d_{123}^{(Sm)}\sigma_{23}}{\mu_0 h}$ $H_2^b = \dfrac{d_{223}^{(m)}\sigma_{23} + d_{213}^{(m)}\sigma_{13}}{\mu_0 h}$ $H_3^b = \dfrac{1}{\mu_0 h}\begin{pmatrix} d_{311}^{(m)}\sigma_{11} + d_{322}^{(m)}\sigma_{22} \\ + d_{333}^{(m)}\sigma_{33} + d_{312}^{(m)}\sigma_{12} \end{pmatrix}$ | $E_1^b = \dfrac{1}{\varepsilon_0 h} d_{113}^{(Se)}\sigma_{13}$ $E_2^b = \dfrac{1}{\varepsilon_0 h} d_{223}^{(Se)}\sigma_{23}$ $E_3^b = \dfrac{1}{\varepsilon_0 h}\begin{pmatrix} d_{311}^{(Se)}\sigma_{11} + d_{322}^{(Se)}\sigma_{22} \\ + d_{333}^{(Se)}\sigma_{33} \end{pmatrix}$ | Size $\eta = h$ 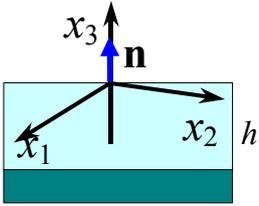 |
| Nanowire of radius $R$, wire axes $\uparrow\uparrow x_1$, local normal $\mathbf{e}_\rho \uparrow\uparrow x_3$ | $H_\rho^b = -\dfrac{2\tau}{\mu_0 R^2}\begin{pmatrix} d_{322}^{(Sm)} + d_{333}^{(Sm)} \\ - 2d_{311}^{(Sm)}\dfrac{s_{12}}{s_{11}} \end{pmatrix}$ | $E_\rho^b = -\dfrac{2\tau}{\varepsilon_0 R^2}\begin{pmatrix} d_{322}^{(Se)} + d_{333}^{(Se)} \\ - 2d_{311}^{(Se)}\dfrac{s_{12}}{s_{11}} \end{pmatrix}$ | Size $\eta = R/2$ 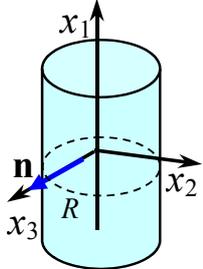 |
| | $\sigma_{33} = \sigma_{22} = -\dfrac{\tau}{R}$, $\sigma_{11} = \dfrac{s_{12}}{s_{11}}\dfrac{2\tau}{R}$, $\sigma_{12} = \sigma_{13} = \sigma_{23} = 0$ $\tau$ is the intrinsic surface stress tensor coefficient | | |
| Nanosphere of radius $R$, local normal $\mathbf{e}_r \uparrow\uparrow x_3$ | $H_r^b = -\dfrac{3\tau}{\mu_0 R^2}\begin{pmatrix} d_{322}^{(Sm)} + d_{333}^{(Sm)} \\ + d_{311}^{(Sm)} \end{pmatrix}$ | $E_r^b = -\dfrac{3\tau}{\varepsilon_0 R^2}\begin{pmatrix} d_{322}^{(Se)} + d_{333}^{(Se)} \\ + d_{311}^{(Se)} \end{pmatrix}$ | Size $\eta = R/3$ 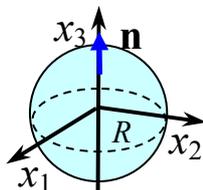 |
| | $\sigma_{33} = \sigma_{22} = \sigma_{11} = -\dfrac{\tau}{R}$, $\sigma_{12} = \sigma_{13} = \sigma_{23} = 0$ | | |
| Thin film of thickness $h$ on a rigid substrate, $u_m$ is misfit surface normal $\uparrow\uparrow x_3$ | $H_3^b = \dfrac{2u_m\left(d_{311}^{(Sm)} + d_{322}^{(Sm)}\right)}{\mu_0 h(s_{11} + s_{12})}$ $\sigma_{11} = \dfrac{u_m}{s_{11}+s_{12}} - \dfrac{E_i\left(s_{11}d_{i11}^{(e)} - s_{12}d_{i22}^{(e)}\right) - H_i\left(s_{11}d_{i11}^{(m)} - s_{12}d_{i22}^{(m)}\right)}{s_{11}^2 - s_{12}^2}$, $\sigma_{22} = \dfrac{u_m}{s_{11}+s_{12}} - \dfrac{E_i\left(s_{11}d_{i22}^{(e)} - s_{12}d_{i11}^{(e)}\right) - H_i\left(s_{11}d_{i22}^{(m)} - s_{12}d_{i11}^{(m)}\right)}{s_{11}^2 - s_{12}^2}$ $\sigma_{12} = \sigma_{31} = \sigma_{32} = \sigma_{33} = 0$ | $E_3^b = \dfrac{2u_m\left(d_{311}^{(Se)} + d_{322}^{(Se)}\right)}{\varepsilon_0 h(s_{11} + s_{12})}$ | *) $\gamma_{ij}^R = \gamma_{ij} + \dfrac{\gamma_{ij}^S}{h} + $ $+ \dfrac{\delta_{i3} d_{311}^{(Se)}\left(d_{j11}^{(Sm)} + d_{j22}^{(Sm)}\right)}{h^2(s_{11} + s_{12})}$ |

*) For a mechanically (partially) clamped system one should also consider the contribution of elastic subsystem into the magnetoelectric coupling $\gamma_{ij}^f$. In the case $\gamma_{ij}^f$ is proportional to the convolution of the surface magnetoelectric and piezoelectric tensors in accordance with formula $\gamma_{ij}^f \sim d_{ilk}^{(Sm)} d_{jlk}^{(Se)}$ (see the third term in the last formulae).

**Acknowledgement**

Supported in part (EAE, ANM) by the Ministry of Science and Education of Ukraine and National Science Foundation (Materials World Network, DMR-0908718).

**Appendix**

For correct phenomenological description of any confined (and in particular nanosized) system the ***surface energy $G_S$*** should be considered, as its contribution in the free energy $G = \int_S g_S d^2r + \int_V g_V d^3r$ increases with the system size decrease ($S$ is the system surface, and $V$ is its volume). The free energy density excess related with the electric and magnetic field components $E_i$ and $H_i$, and the stress tensor $\sigma_{ij}$ has the form [17]:

$$g_S = -\left(d_{ijk}^{(Se)}\sigma_{jk}E_i + d_{ijk}^{(Sm)}\sigma_{jk}H_i + \gamma_{ij}^S H_i E_j\right) \tag{A.1a}$$

$$g_V = -\left(\begin{array}{l}\left(P_{0i} + \dfrac{\varepsilon_0}{2}\chi_{ij}^e E_j\right)E_i + \left(M_{0i} + \dfrac{\mu_0}{2}\chi_{ij}^m H_j\right)H_i \\ + d_{ijk}^{(e)}\sigma_{jk}E_i + d_{ijk}^{(m)}\sigma_{jk}H_i + \gamma_{ij}H_i E_j + \dfrac{s_{jklm}}{2}\sigma_{jk}\sigma_{lm} + ...\end{array}\right). \tag{A.1b}$$

$s_{ijkl}$ is the elastic compliances tensor, $u_{ij}$ are strain tensor components. $P_{0i}$ and $M_{0i}$ are the spontaneous polarization and magnetization vector components. All higher nonlinear and gradient terms were omitted in Eq.(A.1) for the sake of simplicity. In Eq. (A.1) we omitted the gradients of polarization and magnetization, keeping in mind that we are going to consider the region in the vicinity of the surface where the influence of surface is strong. This region corresponds to the shell of the core-shell model proposed earlier [27]. In this model shell and core are two homogeneous regions, shell has close to the surface properties and core has the properties close to the bulk. In accordance with theoretical calculations [26] and ESR experimental data [28] the size of shell lays in the region from several to a few tens of nanometers.

Minimization of the free energy (A.1) in the Gibbs model with respect to the electric, magnetic fields and stress components leads to the equations of state:

$$u_{ij} = -(\partial G/\partial \sigma_{ij}) = s_{ijkl}\sigma_{ij} + \left(d_{kij}^{(e)} + \dfrac{S}{V}d_{kij}^{(Se)}\right)E_k + \left(d_{kij}^{(m)} + \dfrac{S}{V}d_{kij}^{(Sm)}\right)H_k, \tag{A.2a}$$

$$P_i = -(\partial G/\partial E_i) = P_{0i} + \varepsilon_0\chi_{ij}^e E_j + \left(d_{ijk}^{(e)} + \dfrac{S}{V}d_{ijk}^{(Se)}\right)\sigma_{jk} + \left(\gamma_{ij} + \dfrac{S}{V}\gamma_{ij}^S\right)H_i, \tag{A.2b}$$

$$M_i = -(\partial G/\partial H_i) = M_{0i} + \mu_0\chi_{ij}^m H_j + \left(d_{ijk}^{(m)} + \dfrac{S}{V}d_{ijk}^{(Sm)}\right)\sigma_{jk} + \left(\gamma_{ij} + \dfrac{S}{V}\gamma_{ij}^S\right)E_j. \tag{A.2c}$$

In the adopted approximation:

$$G = -V \left( \begin{array}{l} \left(P_{0i} + \dfrac{\varepsilon_0}{2}\chi_{ij}^e E_j\right)E_i + \left(M_{0i} + \dfrac{\mu_0}{2}\chi_{ij}^m H_j\right)H_i + \dfrac{s_{jklm}}{2}\sigma_{jk}\sigma_{lm} + \\ + \left(d_{ijk}^{(e)} + \dfrac{S}{V}d_{ijk}^{(Se)}\right)\sigma_{jk}E_i + \left(d_{ijk}^{(m)} + \dfrac{S}{V}d_{ijk}^{(Sm)}\right)\sigma_{jk}H_i + \left(\gamma_{ij} + \dfrac{S}{V}\gamma_{ij}^S\right)H_i E_j + ... \end{array} \right) \quad (A.3)$$

Here the built-in fields $E_i^b = \dfrac{S}{\varepsilon_0 V}d_{ijk}^{(Se)}\sigma_{jk}$, $H_i^b = \dfrac{S}{\mu_0 V}d_{ijk}^{(Sm)}\sigma_{jk}$ and ME coefficients $\gamma_{ij}^R = \gamma_{ij} + \dfrac{S}{V}\gamma_{ij}^S$ can be naturally introduced and so

$$G_R = -\left(d_{ijk}^{(e)}\sigma_{jk}E_i + \varepsilon_0 E_i^b E_i + d_{ijk}^{(m)}\sigma_{jk}H_i + \mu_0 H_i^b H_i + \gamma_{ij}^R H_i E_j\right)\cdot V. \quad (A.4)$$

**References**

1. A. Sundaresan, R. Bhargavi, N. Rangarajan, U. Siddesh, and C. N. R. Rao, Phys. Rev. B **74**, 161306(R) (2006).

2. Y. Nakal et al., Physica B **284**, 1758 (2000).

3. P. Ayyub, S. Chattopadhyay, R. Pinto, and M. Multani, Phys. Rev. B **57**(10), R5559 (1998).

4. D. Yadlovker and S. Berger, Uniform orientation and size of ferroelectric domains. Phys. Rev. B **71**, 184112 (2005).

5. J. H. Haeni, P. Irvin, W. Chang, R. Uecker, P. Reiche, Y. L. Li, S.B. Choudhury, W. Tian, M. E. Hawley, B. Craigo, A. K. Tagantsev, X. Q. Pan, S. K. Streiffer, L. Q. Chen, S. W. Kirchoefer, J. Levy, and D. G. Schlom. Nature. **430**, 758-761 (2004).

6. A. P. Ferriani, M. Bode, S. Hainze, G. Bihlmouer et al, Phys. Rev. Lett. 94, 087204 (2005)

7. E.A.Eliseev, M.D. Glinchuk. Physica B. 400, 106–113 (2007).

8. A.N. Morozovska, M.D. Glinchuk, and E.A. Eliseev. Phys. Rev. B **76**, 014102 (2007).

9. E. A. Eliseev, M. D. Glinchuk, and A. N. Morozovska. Phys. stat. sol. (b) **244**(10), 3660 (2007).

10. M.D. Glinchuk, A.N. Morozovska, J. Phys.: Condens. Matter. **16**, 3517 (2004).

11. M.D. Glinchuk, A.N. Morozovska, and E.A. Eliseev, J. Appl. Phys. **99**, 114102 (2006)

12. M.D. Glinchuk, A.N. Morozovska, E.A. Eliseev, and R. Blinc. J. Appl. Phys. 105, 084108-1-5 (2009).

13. M.D. Glinchuk, E.A. Eliseev, A.N. Morozovska, R. Blinc. Phys. Rev. B 77, 024106-1-11 (2008).

14. A. Brinkman et al. Nature Mater. **6**, 493–496 (2007).

15. N. Reyren et al. Science 317, 1196–1199 (2007).

16. S Gariglio, N Reyren, A D Caviglia and J-M Triscone, J. Phys.: Condens. Matter 21, 164213 (2009)

17. M.Fiebig, J. Phys. D: Appl. Phys. **38**, R123-R152 (2005).

18. J. F. Scott. Nature Materials, **6**, 256 (2007)

19. B. Ruette, S. Zvyagin, A. P. Pyatakov, A. Bush, J. F. Li, V. I. Belotelov, A. K. Zvezdin, and D. Viehland, Phys. Rev. B **69**, 064114 (2004).

20. J. X. Zhang, Y. L. Li, D. G. Schlom, L. Q. Chen, F. Zavaliche, R. Ramesh, and Q. X. Jia. Appl. Phys. Lett. **90**, 052909 (2007).

21. V.K. Wadhawan. *Introduction to ferroic materials*. Gordon and Breach Science Publishers (2000).

22. E. Roduner. *Nanoscopic Materials. Size-dependent phenomena*. RSC Publishing (2006).

23. Michael G. Cottan. *Linear and nonlinear spin waves in magnetic films and super-lattices*. World Scientific, Singapore (1994).

24. Y. Nakal, Y. Seino, T. Teranishi, M. Miyake, S. Yamada, H. Hori, Physica B, **284**, 1758 (2000).

25. Modern Crystallography. Vol. IV: Physical Properties of Crystals (Springer Series in Solid-State Sciences), ed. L.A. Shuvalov, Springer-Verlag, Berlin, (1988). 538 pages, ISBN: 978-0387115177.

26. M.D. Glinchuk, M.F. Deigen, Surface Science **3**, 243 (1965)

27. A.M. Slipenyuk, I.V. Kondakova, M.D. Glinchuk, V.V. Laguta, Phys. Stat. Solidi C, **4**, 1297 (2007)

28. M.D. Glinchuk, I.V. Kondakova, V.V. Laguta, A.M. Slipenyuk, I.P. Bykov, A.V. Ragulya and V.P. Klimenko. Acta Physica Polonica A, 108(1), 47 (2005).

29. V.I. Marchenko, and A.Ya. Parshin, Zh. Eksp. Teor. Fiz. **79** (1), 257-260 (1980), [Sov. Phys. JETP **52**, 129-132 (1980)].

30. V.A. Shchukin, D. Bimberg, Rev. Mod. Phys. **71**(4), 1125-1171 (1999).